\documentclass[journal]{aastex}
\usepackage{graphicx}

\shorttitle{3D Spectrophotometry of Mrk 493}
\shortauthors{Popovi\'c et al.}

\begin{document}

 \title{3D Spectroscopic Study of the Line Emitting Regions of
      Mrk 493}

\author{
L. \v C. Popovi\'c\altaffilmark{1,2}, A. A. Smirnova\altaffilmark{3}, J.
Kova\v{c}evi\'c\altaffilmark{1,2}, A.V Moiseev\altaffilmark{3}, and V.L.
Afanasiev\altaffilmark{3}}

\altaffiltext{1}{Astronomical Observatory, Volgina 7, 11160 Belgrade,
Serbia}
\altaffiltext{2}{Isaac Newton Institute, Yugoslavia Branch}
\altaffiltext{3}{Special Astrophysical Observatory, Nizhnii Arkhyz,
Karachaevo-Cherkesia,
369167 Russia}

%\date{Accepted 2007  . Received 2007  ; in original form 2007 }
%\pagerange{\pageref{firstpage}--\pageref{lastpage}} \pubyear{2007}

%\maketitle

%\label{firstpage}

\begin{abstract}
We report the results  of 3D spectroscopic observations  of
 Mrk 493 (NLS1 galaxy) with the integral-field spectrograph MPFS of the SAO
RAS 6-m
telescope. The difference in the slope of the optical continuum
emission intensity across the nucleus part { and an extensive continuum emission region}  is detected. { 
The emission in  lines (H$\alpha$, H$\beta$, [OIII], etc.) coincides
 with a composite nuclear region: an AGN plus a circum-nuclear
star-forming ring observed in the HST UV/optical images. The [SII] emission region tends to be up to 1kpc around the center.
 The H$\alpha$ and H$\beta$ could be decomposed into three components (broad $\sim$ 2000 km/s. intermediate $\sim$ 700 km/s and narrow $\sim$ 250 
km/s).
  We found that  width  ($\sim$ 750 km/s) of the Fe II  lines correspond to the intermediate component, that may indicate a non-BLR origin of the 
 Fe II lines, or that a large fraction of the Fe II emission arise in the outher parts of the BLR.  }  The weak broad component detected in the 
H$\alpha$,
H$\beta$ and He$\lambda$4686  may come from the unresolved central
BLR, { but also  partly produced by violent starburst in the circum-nuclear
ring. Moreover,
diagnostic diagrams clearly show presence of the HII regions (not a
Sy 1 nucleus) in the NLR of Mrk 493.}
\end{abstract}

\section{Introduction}

Mrk 493, a narrow-line Seyfert 1 galaxy (NLS1), is known as an
Active Galactic Nucleus with a strong Fe II emission (Osterbrock
and Pogge 1985, Crenshaw et al. 1991). NLS1s have been recognized
as a distinct type of Seyfert nuclei. The optical emission-line
properties of NLS1s can be summarized as  (e.g., Osterbrock \&
Pogge 1985, Komossa 2008): (i) The Balmer lines are only slightly broader than
the forbidden lines such as the [O III] $\lambda$5007 (typically less
than 2000 km s$^{-1}$).
 (ii) The [O III] $\lambda$5007/H$\beta$ intensity
ratio  is smaller than 3.
 (iii) They present the  strong Fe II emission  which are often seen in
  Sy 1s but generally not in Sy 2's.
Moreover, the X-ray spectra of NLS1's are very steep (e.g. Boller
et al. 1996, Leighly 1999ab) and highly variable (Boller et al.
1996; Leighly 1999a; Gliozzi et al. 2007). Also, it was found that
NLS1s have less massive black hole masses than Sy 1s (Mathur 2000, Wang \&
Zhang 2007) and could be young Sy 1s (Botte et al. 2004). Even though NLS1s
have been known for almost more
than 20 years, it is still not clearly understood what NLS1s are in the
context of the current AGN unified model (Nagao et al. 2001).

One of the interesting observational facts connected with NLS1 is
that in their spectra a strong Fe II emission is present. It was
shown by  Boroson \& Green (1992) that  a strong anti-correlation
between the strengths of the ${\rm [OIII]}$ and Fe II line
 intensities exists in the optical  spectra,  where NLS1s are showing the
strongest Fe II
and weakest ${\rm [O III]}$ emission. The observed line widths and
absence of the forbidden emission suggests that the Fe II lines are formed
in the dense broad-line region (BLR), but photo-ionization models
cannot account for all of the Fe II emission. The 'Fe II
discrepancy' remains unsolved, though models which consider
non-radiative heating with an overabundance of iron are promising
(Joly 1993, Collin \& Joly 2000). Also Mathur (2000) proposed
 that the strong Fe II emission observed in NLS1s
 may be connected with their large accretion rates
 and that a collisional ionization origin of Fe II is possible.

Here we report 3D spectroscopic observations of Mrk 493. The aims of
the observations were  (i) to investigate the characteristics of the Fe II
emitting region and possible their connection with other emitting regions
(H$\alpha$, H$\beta$, [OIII], ...)
and (ii) to map the circum-nuclear
emitting gas in this type of AGN. In this paper we accepted a
distance to Mrk 493 of 125.2 Mpc (for
H$_0=75$ km\,s$^{-1}$\,Mpc$^{-1}$), where 1$''$ on the sky
corresponds to 610 pc at the galaxy distance.

\section{Observations and data reduction}

The observational data were obtained at the prime focus of the
Russian 6-m telescope of the Special Astrophysical Observatory
(SAO). The central region of the galaxy was observed with the
MultiPupil Fiber Spectrograph (MPFS). MPFS (Afanasiev et al. 2001)
takes simultaneous spectra from 256 spatial elements (constructed in
the shape of square lenses) that form on the sky an array of
$16\times16$ elements with the angular size 1 arcsec/element.
  Bundle of  17 fibers placed on the distance about 3.5 arcmin from the lens
 array provides the night-sky background spectra simultaneously with
object's exposition.
 We have observed Mrk 493 twice in years 2004 and 2007, in the low- and high-
resolution spectral modes (see Tab.~\ref{speclog}).

The spectral ranges included numerous emission lines of ionized gas
([OI]$\lambda6300$, [OIII]$\lambda\lambda4959,5007$, [NII]$\lambda\lambda6548,6583$,
[SII]$\lambda\lambda6717,6731$, the Balmer H$\alpha$ and H$\beta$ lines).  The
spectra were reduced using the
IDL-based software developed at the SAO RAS, and the data reduction
sequence is briefly described in Moiseev et al. (2004). Namely, the  primary
reduction included bias subtraction, flat-fielding, cosmic-ray hits removal,
extraction of individual spectra from the CCD frames, and their wavelength
calibration using a spectrum of a He-Ne-Ar lamp. Subsequently, we subtracted
the night-sky spectrum from the galaxy spectra.   We corrected the effect of
differential atmospheric refraction (offset shifting of galactic images at
different wavelengths) by using the algorithm similar to described by Arribas
et al. (1999). The data reduction result in a data cube in which a spectrum
corresponds to each image {`spaxel'}\footnote{ the spatial element of a 3D spectra} in two-dimensional field $16\times16$ arcsec.
The seeing values shown in Table.~\ref{speclog} were  estimated from the MPFS
observations of stars right before the object observations. Here we fitted by  2D Moffat function the stars  images in their data cubes.

Additionally we observed broad-band  direct images of the galaxy in
the filters B,V and R$_C$ with the focal reducer SCORPIO (Afanasiev
\& Moiseev 2005). The seeing was 1.3 arcsec, {the total expositions were
400 sec in B and  360 sec in V and R$_C$ bands, respectively.} These images seem
slightly deeper than POSS and SDSS archival data, we confidently
detected  outer low-brightness grand-design structure in this barred
galaxy (Fig.~\ref{fig_im}).

\section{Results and discussion}

\subsection{The continuum and line fluxes distribution across Mrk 493
nucleus}

First, we explore the brightness distribution across Mrk 493 nucleus
in different wavelength bands. In Fig~\ref{fig_2D}  we presented the
maps of emission lines.  {Each pixel on these maps corresponds to the flux
obtained by Gaussian fitting of the lines or simple by integration around 6200 \AA\  in the continuum.}
The maps of the [OIII]$\lambda\lambda5007$ and
[SII]$\lambda\lambda6717,6731$ were constructed using a single-Gaussian
fitting, while the maps in the H$\beta$ and  H$\alpha$ regions {are separately constructed for
the narrow as well as for the broad Gaussian components of these lines}. Also, we
made a map of the Fe II features (left and right from H$\beta$+[OIII] lines),
integrating the all emission (after subtracting the continuum) in the wavelength range
from 4500 -- 4800 \AA\ (blue shelf) and 5200 -- 5500 \AA\ (red
shelf).  The maximum of the Fe II emission corresponds to the optical center of Mrk 493 (i.e. the
continuum emission).

In order to  investigate  the  spatial extension of the
emission-line regions we fitted the images of the star-like nucleus
with 2D Moffat function. { We assumed that the FWHM of the 2D Moffat function represents the dimension of the regions. The results of the fitting are  shown in
Fig. ~\ref{fig_2D}. We should note here that the sizes (FWHM) of
the nuclear emission line region (except [SII] lines) are almost the
same or slightly larger ($2''$) than was the seeing during the observations ($1.7\pm0.2''$).
Therefore, the dimension of the  Fe II, [OIII] and the narrow and broad
Balmer line emitting regions seem to be practically the same. It might be
caused by    spatial resolution of the instrument,  but in any case there is a compact
[OIII] region,   smaller than continuum and [SII] one (see Fig. 2).
The [SII] emission region tends to be up to 1 kpc   around the center.}

Also, we found that the continuum slope and the H$\alpha$/H$\beta$ line flux ratio
is changing across
the circum-nuclear region (an example is shown in Fig. 3). It  indicates
(as well as the dimension of the continuum source, see Fig. 2) that
the contribution to the continuum is coming not only from the
nucleus, but also from the circum-nuclear region. { The blue bump
(see Fig. 3) in the continuum flux may be caused by the UV-radiation from an
accretion disk or by the violent star-formation in the circum-nuclear
ring.} Note that a similar effect can be caused by the different dust absorption
(inartistic reddening) across the central part.

\subsection{Fe II emission region in Mrk 493}

As it can be seen in Table 1, { 3D spectra of Mrk 493 were observed in} two
epochs, in 2004 and 2007. Consequently, first we were looking for the line/continuum
changes between these epochs. To do that, we found a summary spectra for both epochs
around the central part. To avoid errors caused by procedure of spectra elaboration (flux
calibration, different number of spaxels taken in summary spectra in different epochs, etc.)
we normalized both spectra on the maximal intensity of the H$\beta$ line.
 In Fig. 4, we presented the normalized spectra to the maximal
intensity of the H$\beta$ line. As it can be seen in Fig. 4, there is no significant
difference between the spectra (in the lines and continuum) observed in these two epochs (a three-year period).
 Only, a
slight difference can be noted in the He II$\lambda$4686\AA\ wavelength range (see Fig. 4, down),
that is probably caused by changing in this line. A marginal difference between
the spectra (see Fig. 4 down, for $\lambda<$4750\AA ) is
  caused by different spectral resolution (see Table 1). Note here that Klimek
et al. (2004)
found a slight long-term variation in V-band of Mrk 493 ($\sim\pm$0.15$^m$, see their Fig. 18).

In order to find the parameters of the Fe II line emitting region
and its possible correlation with the H$\beta$ line emitting one, we
use the high resolution spectra. We fitted the Fe II and H$\beta$
lines in the nucleus region in order to compare  the kinematic
parameters of the  Fe II and  H$\beta$ line emitting regions
(Fig. \ref{fig_hb}. Also, we fitted  the
H$\alpha$+[NII]  wavelength band (Fig. 6) in the low-resolution spectral
data cube in order to compare kinematics of the H$\beta$ and
H$\alpha$ emitting regions.

There are several ways to fit Fe II template (red and blue shelf,
see e.g. Boroson \& Green 1992, Popovi\'c et al.
2004, Veron et al. 2006, etc.). { We use a template of 53 the most 
intensive Fe II lines in the
spectral  interval from 4400 \AA\ to 5400 \AA\ from five types of
transitions  with  $3d^6(^3F2)4s{\ }^4F$, $3d^54s^2 {\ }^6S $, $3d^6(^3G)4s{\ }^4G$,
 $3p^63d^7 {\ }^2D $ and $3d^6(^5D)4s{\ }^6D$
lower level configurations. Each of the 53 lines is
assumed to have Gaussian profile.}

To constrain the
number of fitting parameters, we assumed:

 (i) all Fe II lines are
coming from the same emitting region, therefore the widths ($W$) and
shifts ($d$) of the Gaussians representing the lines are the same;

(ii)
the line intensities within one transition defined by low level
configuration have been connected with the relation (Kova\v cevi\'c et al. 2008):
{
\begin{equation}
 \frac{I_1}{I_2}\approx{(\frac{\lambda_2}{\lambda_1})}^3\frac{f_1}{f_2}\cdot\frac{g_1}{g_2}
\end{equation}
where $I_1$ and $I_2$ are  intensities of the lines with the same lower level of a transition, $\lambda_1$ and $\lambda_2$ are  line transition wavelengths, $g_1$ and $g_2$ are  corresponding statistical weights, and $f_1$ and $f_2$  the oscillator strengths, $E_1$ and $E_2$ are  energies of the upper level of transitions.}

 (iii) the
intensity ratio of the [OIII]$\lambda\lambda$4959,5007 \AA \ lines was
fixed as 1:3 (see Dimitrijevi\'c et al. 2007), and we also assumed
that all narrow lines ([OIII] and narrow H$\beta$) have the same
width and shift.

(iv){  In summary, we fitted the wavelength region FeII+H$\beta$+[OIII] with a sum
of Gaussian functions (Fig. 5A), from which 53 represents the Fe II template (heaving the same widths and shifts, Fig. 5C).} To find the best fit we use  $\chi^2$
minimization routine.

Also, we used low resolution spectra to fit (with the same sum of Gaussian functions)
the   H$\alpha$ and H$\beta$ lines. We were able
to find a good  and consistent fit (having the similar parameters for corresponding Gaussian function
 in both lines)    using three-Gaussian fitting (see Figs. 5 and 6). Two Gaussian { model
 for broad H$\alpha$ and H$\beta$ cannot properly fit the lines profiles}, since there are  
far wings in these lines (especially
in H$\alpha$, see Fig. 6). Moreover, a F-test shows that the 
three Gaussian { assumption is statistically better than the  two Gaussian one.}
  Note here that often three regions in AGN have
been considered (see e.g. Sulentic et al. 2000). Consequently, we fitted H$\alpha$
and H$\beta$ with three Gaussian functions, aiming that the parameters of Gaussian for
the same spectrum (obtained from one spaxel) have similar values for both lines (fitting the spectra
observed in 2007). Note here that recently Hu et al. (2008a) reported that the
BLR of AGN is composed from two emitting regions.

We were able to { fit the emission lines from 16 and  21 spaxels of high (observed in 2004) and low
(observed in 2007) resolution spectra, respectively. In both cases, we  selected only spectra around the center where in each of them
 the  H$\beta$ or/and H$\alpha$  can be decomposed
into three components (broad - W$_B$, intermediate W$_I$ and narrow
$W_{NLR}$)  and  where the signal-to-noese ratio was enough good that the far wings are well defined. The parameters of the best fit of spectra from each spaxel are given in Tables 2-4. In Table 2-3 the parameters of the best fit for 16 high resolution spectra are presented. As it can be seen from Tables 2-3, the obtained parameter values for different spaxels are consistent, and  differences between a parameter from different spaxels is probably caused by the accuracy of the fitting procedure. The same is in the case of the low resolution spectra (Table 4).    As one can see from Tables 2-3, the averaged values of widths of the H$\beta$ line components are: 
W$_B$=(2380$\pm$ 200) km/s, W$_I=$(790$\pm$80) km/s and 
$W_{NLR}=$(250$\pm$40) km/s\footnote{$W$ is taken from
Gaussian function $exp(-(\delta\lambda-d/W)^2)$
and it is connected with velocity dispersion ($\sigma$) and Full Width at
Half Maximum (FWHM) as $W=\sqrt{2}\sigma=FWHM/2\sqrt{ln(2)}$} ($W_{NLR}$ is assumed to be the same as the narrow [OIII] line width).}
 The broadest component is weak and
it is interesting that it is slightly more intensive than the
He$\lambda$4686 line (Gaussian represented with solid line left from
the H$\beta$ in Fig. 5B). In all spectra the broadest component shows { the blue shift of
 -(429$\pm$133) km/s} with respect to the narrow [OIII] lines, while
the intermediate component  
 shows a slight redshift; { in average +(156$\pm$47) km/s,  that is in
the frame of fitting errors}.  Also, there is no significant difference
between the Gaussian parameters across the nucleus part, the
difference is probably caused by accuracy of the fitting with three
Gaussian functions. It means that we were not able to find significant differences
in gas kinematics across the nuclear part.

{ Also, we fitted the H$\beta$ and  H$\alpha$ lines from the low resolution spectra (observed in 2007, 21 spaxels). The parameters of the best fit for H$\alpha$ are given in Table 4. The obtained parameters for H$\beta$ from this spectra are similar to the parameters shown in Tables 2 and 3 (for high resolution spectra), but showing systematically  slight larger widths caused by lower resolution. As one can see from Table 4, the H$\alpha$ ($W_{BLR}\sim 1800$ km/s and $W_{ILR}\sim 650$ km/s) has a smaller width and shift than H$\beta$. The difference in the shift and width between the broad and intermediate H$\alpha$ and H$\beta$ may be caused partly by fitting procedure and partly by different spectral resolution.}

The  Fe II line { widths from different spaxels} indicate the velocity gas in the Fe II
emission region around W$_{FeII}\approx $800 km/s { (an averaged value 760$\pm$50 km/s, see Table 2)}, that corresponds to the random velocity in the
intermediate region  (as well as the shift, { see Tables 2 - 4}). { This fact suggests that the Fe II line could be} originated in
an intermediate line region as it was mentioned in Popovi\'c et
al. (2004). { Moreover, recently Kuehn et al. (2008) suggested that  the optical Fe II emission of Ark 120 does not come from a photoionization-powered region similar in size to the H$\beta$ (broad component) emitting region, i.e. they conclude that Fe II  emission come from a photoionized region several times larger than the H$\beta$ one, or from gas heated by some other mechanism (not by photoinozation). Moreover, Hu et al. (2008b) found the the line width of Fe II is significantly narrower than that of the broad component of H$\beta$ in a sample of 4037 AGN. All of the results mentioned above as well as our fitting results for the Fe II lines suggest that  a large fraction of the Fe II emission arise in the outer parts of the BLR, i.e.  Fe II emission may be originated in the intermediate-line region, which may be the transition from the torus to the BLR (or an accretion disk, see  e.g. Popovi\'c et al. 2004).} 
{ From this observations} is hard to say if the portion of the Fe II emission is formed
  in the BLR, since the BLR component is too weak even in the Balmer lines
(Figs. 5 and 6) { and contributes around 25-40\% to the total line flux of Balmer lines (see Tables 3 and 4). The Fe II emission is strong and the flux ratio of  the Fe II emission (integrated in the interval 4400 \AA\ to 5500 \AA ) and H$\beta$-total line flux is 0.52$\pm$0.08.}

The Fe II emission spectra formation from AGN is still
poorly understood (Joly 1993; Hamann \& Ferland 1999, Collin \&
Joly 2000). Typically, photoionization cloud models for the BLR fail to
account for the observed strength of the Fe II emission. The
mechanism that is responsible for Fe II origin is supposed to be
the collisional excitation (see e.g. Sigut \& Pradhan 2003), i.e.
inelastic collisions with electrons excite the odd parity levels
near 5 eV which then decay into the optical and UV lines. This
mechanism is efficient whenever the gas temperature is above 7000
K, temperatures which are generally found in photoionized models of
the BLR. Excitation is irrespective of the local optical depth in
the Fe II lines, and thus this mechanism does not suffer 
limitations of the continuum fluorescence. It is generally believed
that collisional excitation is responsible for the bulk of the Fe
II emission. Also, Collin \& Joly (2000) proposed that Fe II
emission can be explained by a non-radiative heating mechanism, as
e.g. shocks, with an overabundance of iron. Comparison of AGN with
other objects emitting strong Fe II lines is in favor of  the
presence of strong outflows and shocks. { It is interesting here
that we could not register significant outflow in the Fe II emitting
region of Mrk 493, but there is a systematic blueshift of the broad H$\alpha$ and H$\beta$
component. This asymmetry can indicate some kind of the outflow. Note here that Hu et al. (2008b) found that majority of quasars (from a sample of 4037 quasars) show redshifted Fe II emission toward to red around 400 km/s.}

\subsection{The source of ionization of the NLR of Mrk 493}

In order to identify the nature of the gas ionization source in Mrk
493, we constructed diagnostic diagrams using the { narrow} emission-line
intensity ratios. In Fig.~\ref{fig_diagram} we present [O
III]/H$\beta$ vs. [N II]/H$\alpha$ and [O III]/H$\beta$ vs. [S
II]/H$\alpha$. According to Veilleux \& Osterbrock (1987) we
separated the regions corresponding to the ionization by AGN, young
OB-stars (H II regions) and shock waves (LINERS) on the
diagrams.

It was
surprising, that all points are only in the H II region, implying
that the main source of ionization should be young hot stars from
violent star formation in the center of Mrk 493.

 From these diagrams we can conclude that the low ionization lines are primarily excited by star formation.
No any contribution to line ionization from AGN continuum is seen in  Fig.~\ref{fig_diagram}.

  In the analysis is not taken into account a possible absorption in the
 H$\alpha$ and H$\beta$ lines. In any case the stellar absorption reduction
would increase of the H$\alpha$
 and H$\beta$ fluxes, consequently  the  [OIII]/H$\beta$ ratio would be lower;
and points presented in Fig. 7
would be deeper in { the H II section of the graph}. Therefore, we can point out that
star forming process is probably
a dominant source of ionization in the central part of Mrk 493.

\subsection{The circum-nuclear ring}

As it can be seen in Fig. 1, in the central part of the Mrk 493, a
circum-nuclear ring of star-formation is present, {already mentioned
by Deo et al. (2006), see also new HST/ACS observations in the UV
domain (Mu\~{n}oz-Mar\'{i}ne et al. 2007). 
As it can be seen in Fig. 1, this is a prototypical galaxy with a strong large-scale bar with leading edge
dust lanes feeding a central nuclear ring and a grand-design spiral
toward the center. From the HST images, it  can be seen a presence
of multiple dust spiral arms outside the nuclear ring. {Note here that the radius of the ring
(0.6-0.8 arcsec corresponding 350-500 pc) is under the spatial resolution of our MPFS data, but there is also an extensive continuum emission (elliptical shape, see Fig. 2) that is probably coming from emission gas around this ring.}

{All properties of the ring (its size, location inside the strong
bar, morphology of dust lanes) indicate the resonance origin of this
structure. It seems to be a nuclear ring formed  near the Inner
Lindblad Resonance (ILRe) of a large-scale bar (see Buta \& Combes,
1996 for details). According to the common point of view a
large-scale bar efficiently drives gas from the outer disk into the
inner region where the gas flow stops near the ILRe where the
powerful starburst are commonly observed (for instance, see review
by  Jogee 2006). }

{We roughly estimated the total rate of starformation in the
circum-nuclear region using the formula from Kennicutt (1998) which connects
the SFR with the H$\alpha$  luminosity ($\rm L_{H\alpha}$). From MPFS data we
estimated the $\rm L_{H\alpha}=6.7\times10^{41}$ erg\,s$^{-1}$ in the aperture
radius of 2.5 arcsec. This luminosity corresponds to  SFR=$5 M_\odot/yr$. Of
course, the total $\rm L_{H\alpha}$ includes also the nuclear AGN
contribution  and therefore the SFR value is overestimated. { In Table 4, we give the estimated contribution of the broad H$\alpha$ component (and intermediate one) to the total flux of H$\alpha$ (columns 7 and 8) and as it can be seen from  Table 4, around 40\% of H$\alpha$ is emitted in the narrow H$\alpha$ line. Consequently, it can indicate  the SFR=$2M_\odot/yr$}, that  is unusually large for such compact region (less than one
kiloparsec in the diameter).  }

{ In a such system it is interesting to see the spectra further from the centra. Even we have very weak spectra outside 3"x3", we were able to see the weak H$\alpha$ and H$\beta$ lines in the spaxels located at 4" from the center (Fig. 8, upper panel). In Fig. 8 (upper panel) we showed the spectrum extracted from three spaxels: at the center ((0,0) - top), 2" far from the center ((0,2) - middle) and 4" far from the center ((0,4) - down). To compare these three spectrum,  the last two spectra  are magnified by 3 and 27 times, respectively. Although the spectra at (0,4) is very noisy we were able to compare the H$\alpha$ line profile from this spectrum (solid line in Fig. 8, panel bottom) with the two closer to the center (dashed and dashed-doted line for spectra from 0" and 2", respectively -- see Fig. 8 panel down). Also, we normalized the H$\alpha$ lines to one. As it can be seen from Fig. 8, the line profiles of H$\alpha$ from 0" and 2" are almost identical, while H$\alpha$ from (0,4) tends to have stronger [NII] emission, but it should be taken with caution since small signal-to-noise ratio in this spectrum. It is interesting that the H$\alpha$ line profiles  are similar, i.e. that broad wings are present in the H$\alpha$ line at 4"  from the center. It may also indicate that some other mechanism can contribute to the origin of the broad Balmer line component in Mrk 493 (non-AGN component)}

{ Since   the stellar ring is located in the central part of the line emission regions one can speculate that the ring may be in principle star forming region.} It is interesting that the radiation of young
stars from the ring probably represents the source of ionization of
the line emitting regions in Mrk 493. Moreover, the strong Fe II
emission and weak [OIII] one, can be explained with a model that
contains massive starburst (SB) plus an AGN (see e.g. L\'ipari \&
Terlevich 2006, etc.). The SB+AGN can lead to large scale expanding
supergiant shells (see e.g. L\'ipari et al. 1994, 
Lawrence et al. 1997, Canalizo et al.
1998, L\'ipari \& Terlevich 2006). In the case of Mrk 493,   it
might be that the Fe II lines are also formed in the gas located in (or
around) the ring. The question is why the Fe II lines are broad
around 800 km/s? {Note here that bloated stars extended envelopes
are similar, in many ways, to the low density gas. Close to the
center they will show typical BLR spectrum and further away a
typical NLR spectrum (Netzer 2006). Since violent star formation
processes are in the central part of Mrk 493, and our analysis of
ionizing diagrams shows that the ionization source is only thermal,
we can not exclude a possibility that the star formation process is
the main mechanism of radiation in the center of this galaxy.}
 Moreover, the active star formation-regions have been
found in
 Seyfert galaxies (see e.g. Davies et al. 2004, 2007; Mu\~{n}oz-Mar\'{i}ne et
al. 2007, and references therein). E.g. Davies et al.
(2007) { found
that in the case of nine Seyfert galaxies, on kpc scale, the
stellar luminosity is comparable with AGN one.}

\section{Conclusion}

In this paper we report the 3D spectrophotometric observations of the
Mrk 493. From our investigation we can give the following
conclusions:

(i) {The continuum and [SII] line emitting regions in Mrk 493 seem to be
 very extensive ([SII] up to 1 kpc, and continuum $>$1kpc), and they are larger than
the  [OIII], Fe II and Balmer emission line regions.}

 (ii)  It seems that the strong Fe II emission is not coming from the BLR
 ($W_{BLR}\sim$2000 km/s),
 but from a region with velocity dispersion around 800 km/s. { This is in an agreement 
with  
some earlier investigation given by Popovi\'c et al. (2004) and recent ones given by Kuehn et al. (2008) and Hu et al. (2008) that a large fraction of the Fe II emission may come from  an ILR.}

(iii)  {  There is no   changes of the continuum and emission  lines (instead He
II$\lambda$4686)  between two observations (in the three-year period), that may also indicate
 non-BLR origin of  lines. Otherwise,
the BLR in Mrk   493 should be a): very extensive $\sim$ 1pc and/or b) very stable.}

(iv) The source of ionization of the narrow Balmer and [OIII] line
emitting regions in Mrk 493 seems to not  be the nucleus,
but the star-forming processes;

(v) { The position  of the line and continuum regions coincides 
with a composite nuclear region: an AGN plus a circum-nuclear
star-forming ring observed in the HST UV/optical images, but it seems that the emitting source of the 
narrow line and continuum is probably the circum-nuclear
star-forming ring, but here remains a question about the mechanism
of the Fe II production in a such  ring.}

\section*{Acknowledgments}

This work is  based on the observational data obtained with the
6-m telescope of the Special Astrophysical Observatory of the
Russian Academy of Sciences funded by the Ministry of Science of
the Russian Federation (registration number 01-43) and from the
data archive of the NASA/ESA Hubble Space Telescope at the Space
Telescope Science Institute. This work was supported by the
Ministry of Science  of Serbia through the project (14002)
"Astrophysical Spectroscopy of Extragalactic Objects". Also, the
work has been financed by the Russian Foundation for Basic
Research (project no.~ 06--02--16825). A.V.M. acknowledges a grant from the President
of the Russian Federation. { We would like to thank to the referee for very useful comments.}

\clearpage

\begin{figure}
\begin{center}
\includegraphics[width=0.5\textwidth]{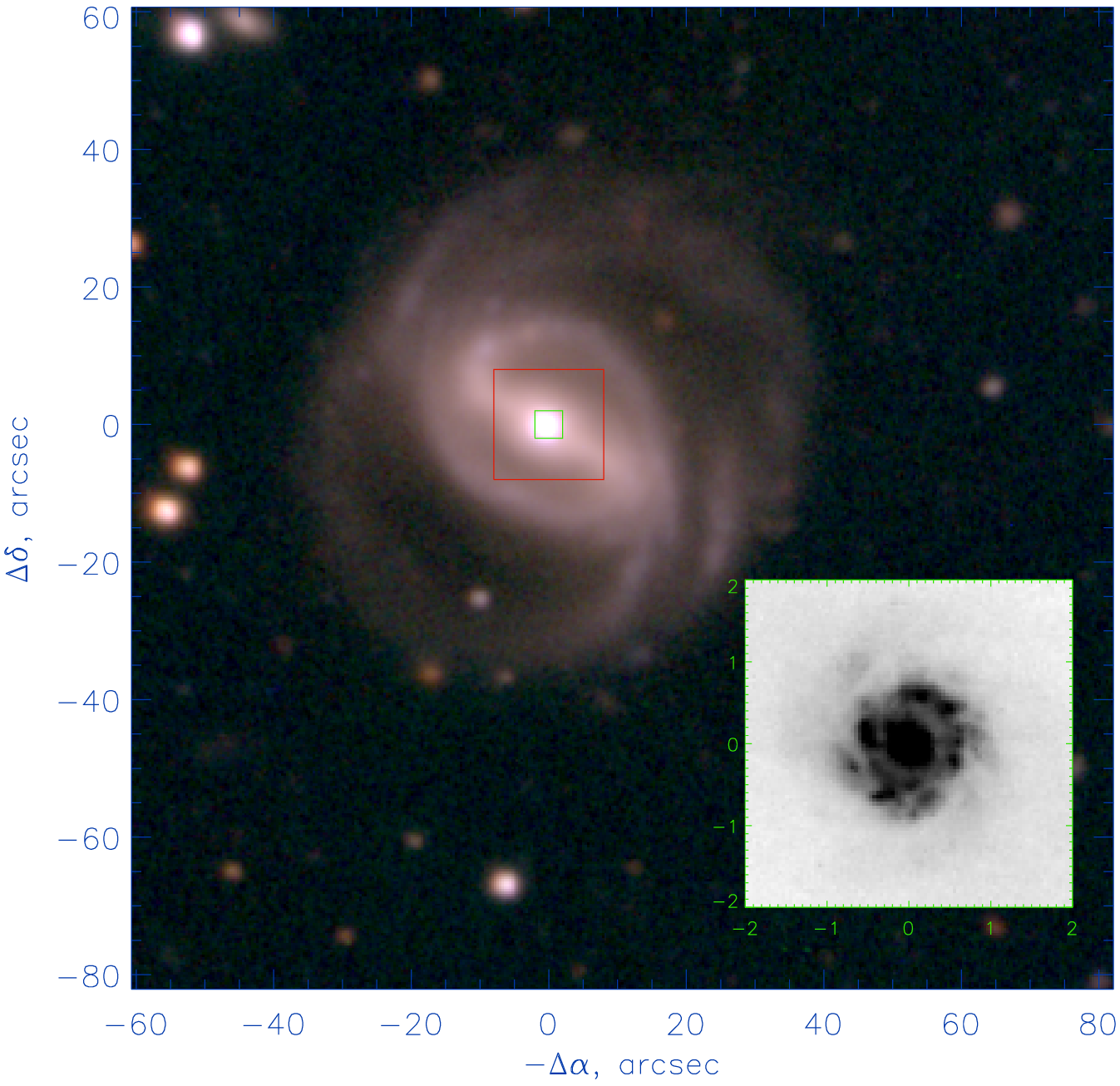}
\caption{The composite BVR$_C$ color image obtained with SCORPIO at
6-m telescope. The large red box shows the region observed with
MPFS. The small green box marks the field shown  in the bottom
right corner. It is the HST WFPC2 image in the optical band with
F606W filter.} \label{fig_im}
\end{center}
\end{figure}

\begin{figure*}
\begin{center}
(a)
\includegraphics[width=1\textwidth]{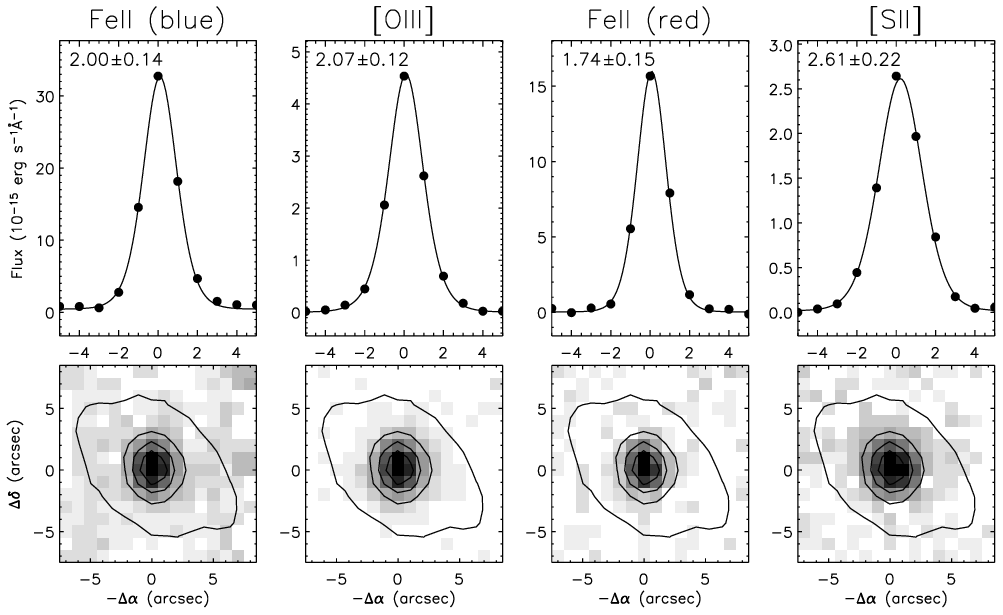}
(b)
\includegraphics[width=1\textwidth]{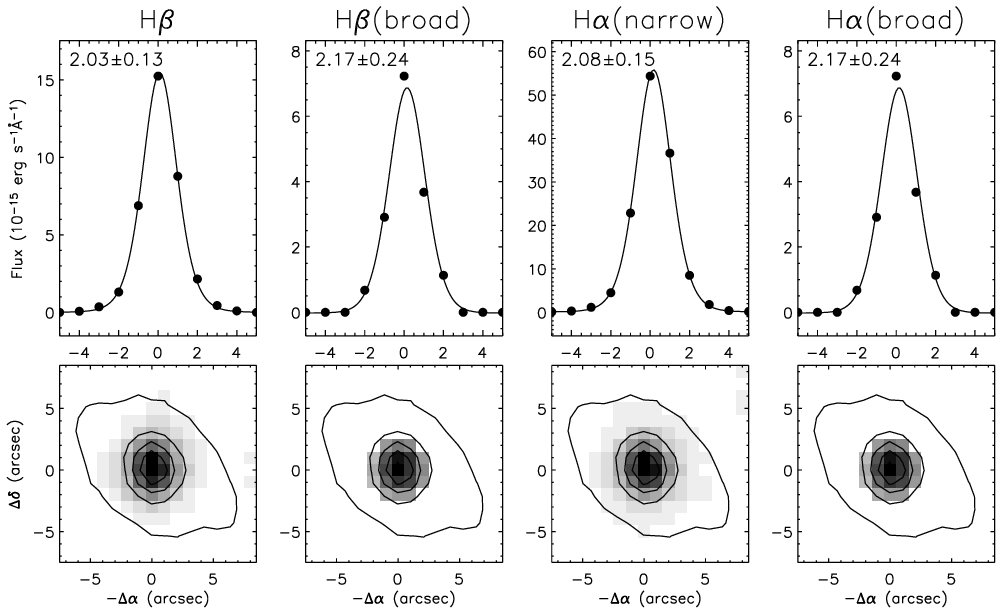}
\caption{The  circum-nuclear region of Mrk 493 observed with MPFS.
{The maps in the [OIII]$\lambda5007$, [SII]$\lambda6717,6731$ and Fe II lines (a)
and in narrow and broad components of the Balmer lines (b) }
The contours represent the continuum measured near $\lambda6200$\AA .
Top raw show the cross-sections of the nuclear emission  along the
horizontal axis. The points are observational data, while the lines
are approximation with the 2D-Moffat function. The FWHMs of these
profiles are given in the upper left corner on each plot.}
\label{fig_2D}
\end{center}
\end{figure*}

\begin{figure}
\begin{center}
\includegraphics[width=0.5\textwidth,angle=-90]{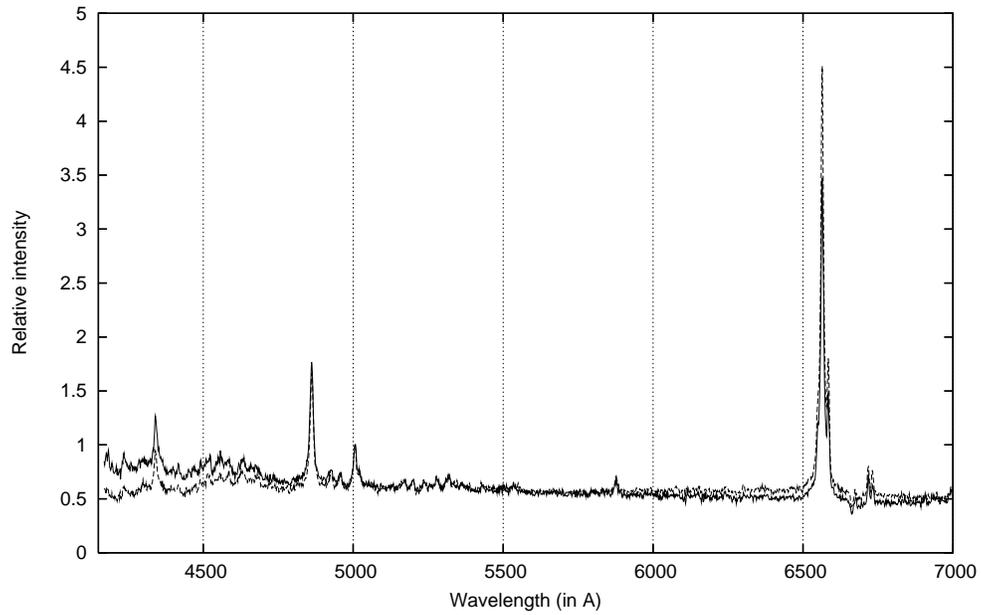}
\caption{ Comparison of spectra from different spaxels, with coordinates relative the nucleus
($\Delta\alpha=-1'',\Delta\delta=-2''$)  and ($\Delta\alpha=+2'',\Delta\delta=+1''$ ),
of Mrk 493. The spectra are normalized to the maximum of the H$\beta$ line. } \label{fig1}
\end{center}
\end{figure}

\begin{figure}
\begin{center}
\includegraphics[width=.5\textwidth,angle=0]{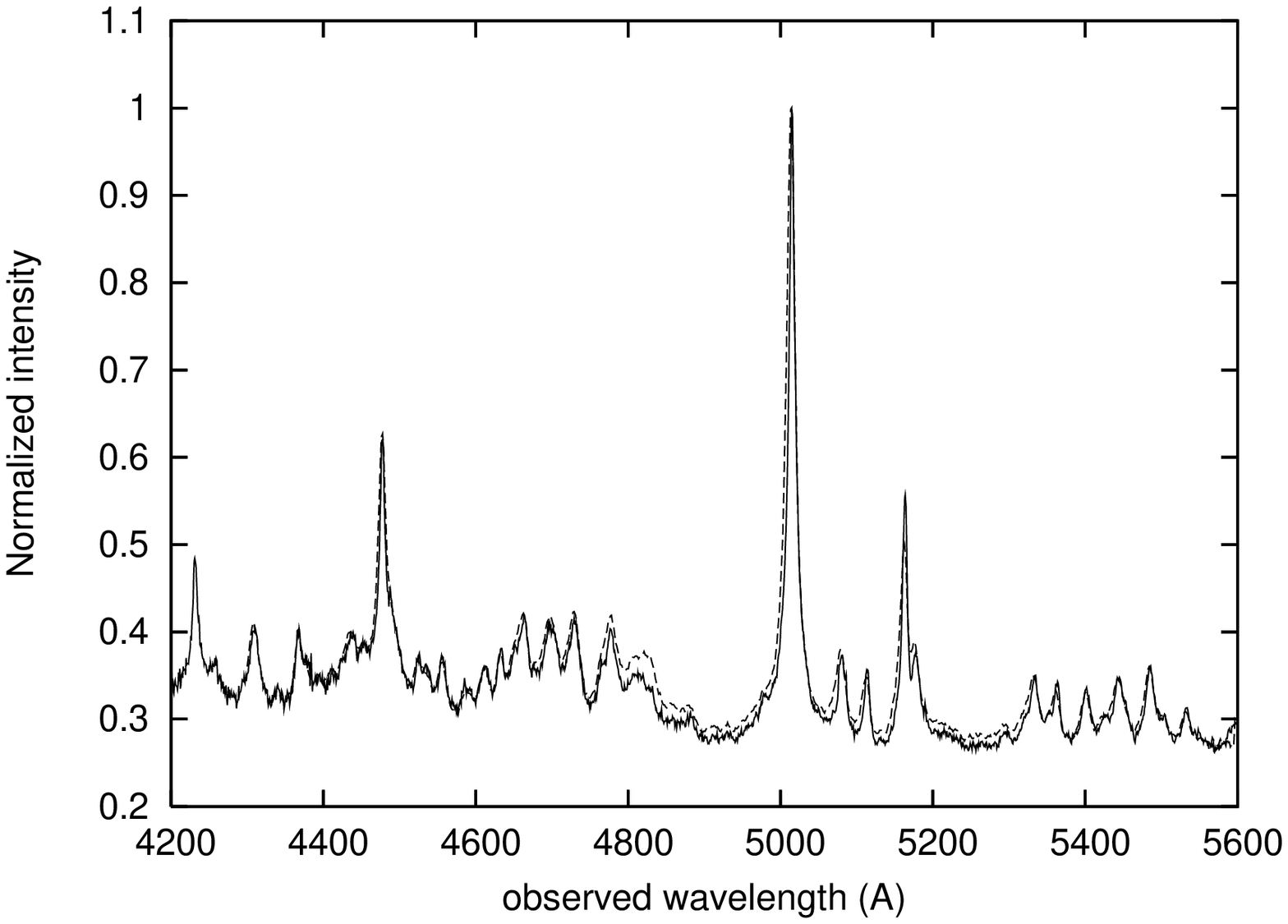}
\includegraphics[width=.5\textwidth,angle=0]{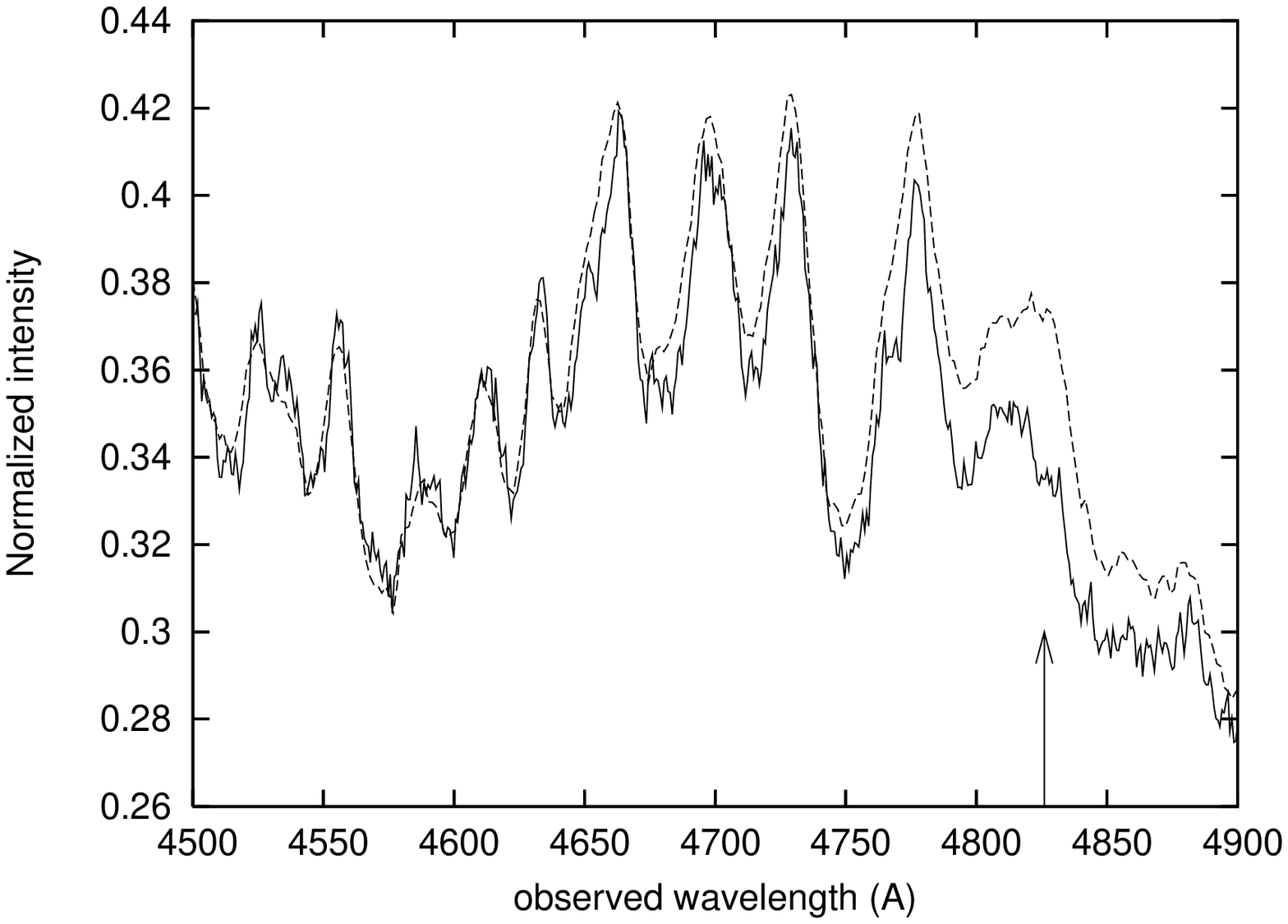}
\caption{  Panel up: The Mrk 493 spectra in the H$\beta$+Fe II  wavelength
band normalized
to the maximum   of the H$\beta$ intensity, observed in 2004 (solid line)
and
2006 (dashed line).
Panel down: The blue Fe II  template observed in 2004 (solid line) and
2006
(dashed line), the arrow
shows the position of the  He II$\lambda$4686 line in observed wavelength
scale. } \label{fig_com}
\end{center}
\end{figure}

\begin{figure*}
\begin{center}
\includegraphics[width=.5\textwidth,angle=-90]{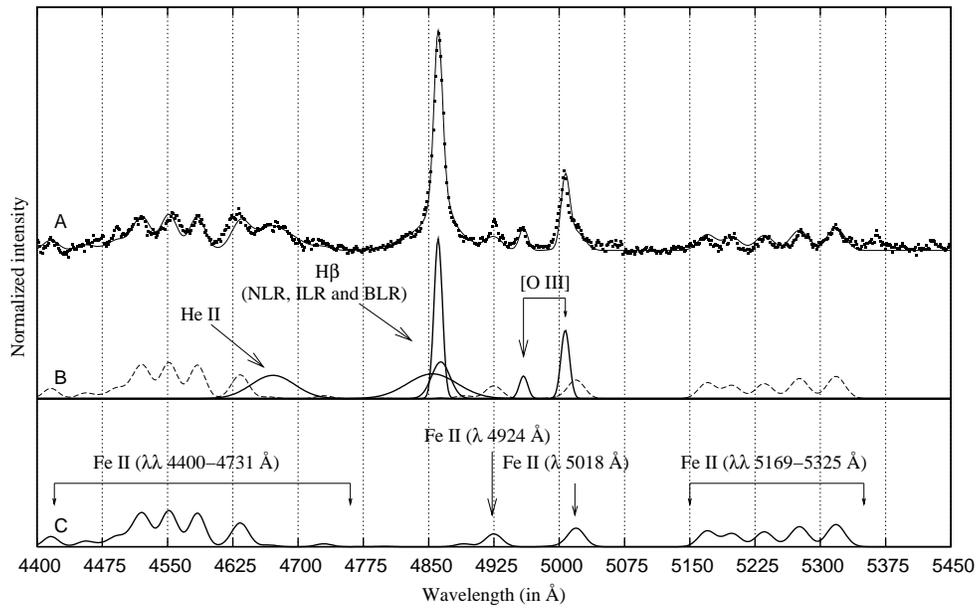}
\caption{ The decomposition of the observed H$\beta$ spectral region
(dots)
 using the $\chi^2$  best fit minimization (C). The Gaussian
functions denoted with
solid lines in panel B represent the components of He II, H$\beta$ and
[OIII]. The Fe II template obtained from the best fit (53 Gaussian
functions) is presented in panel C. } \label{fig_hb}
\end{center}
\end{figure*}

\begin{figure}
\begin{center}
\includegraphics[width=.5\textwidth,angle=0]{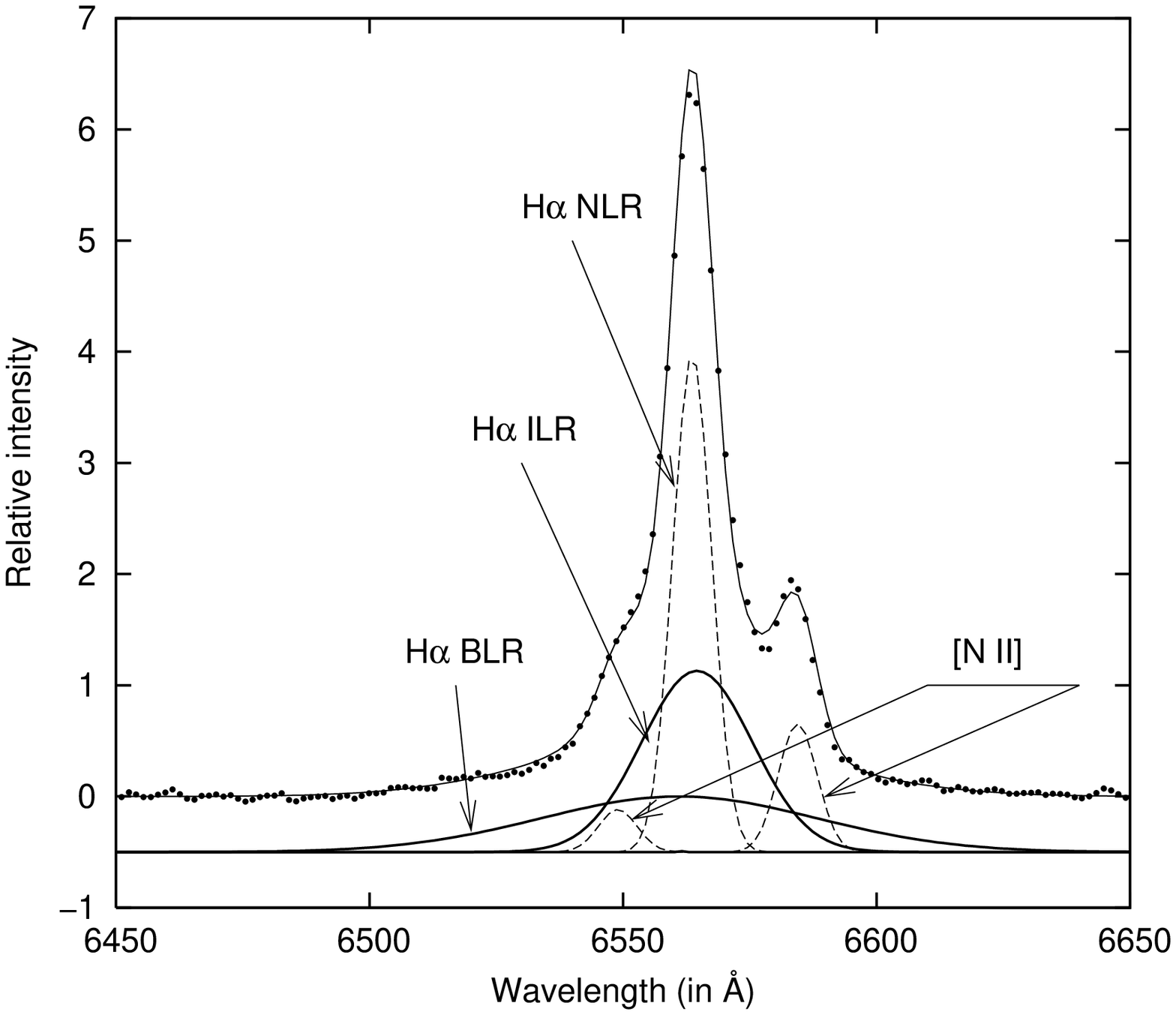}
\includegraphics[width=.5\textwidth,angle=0]{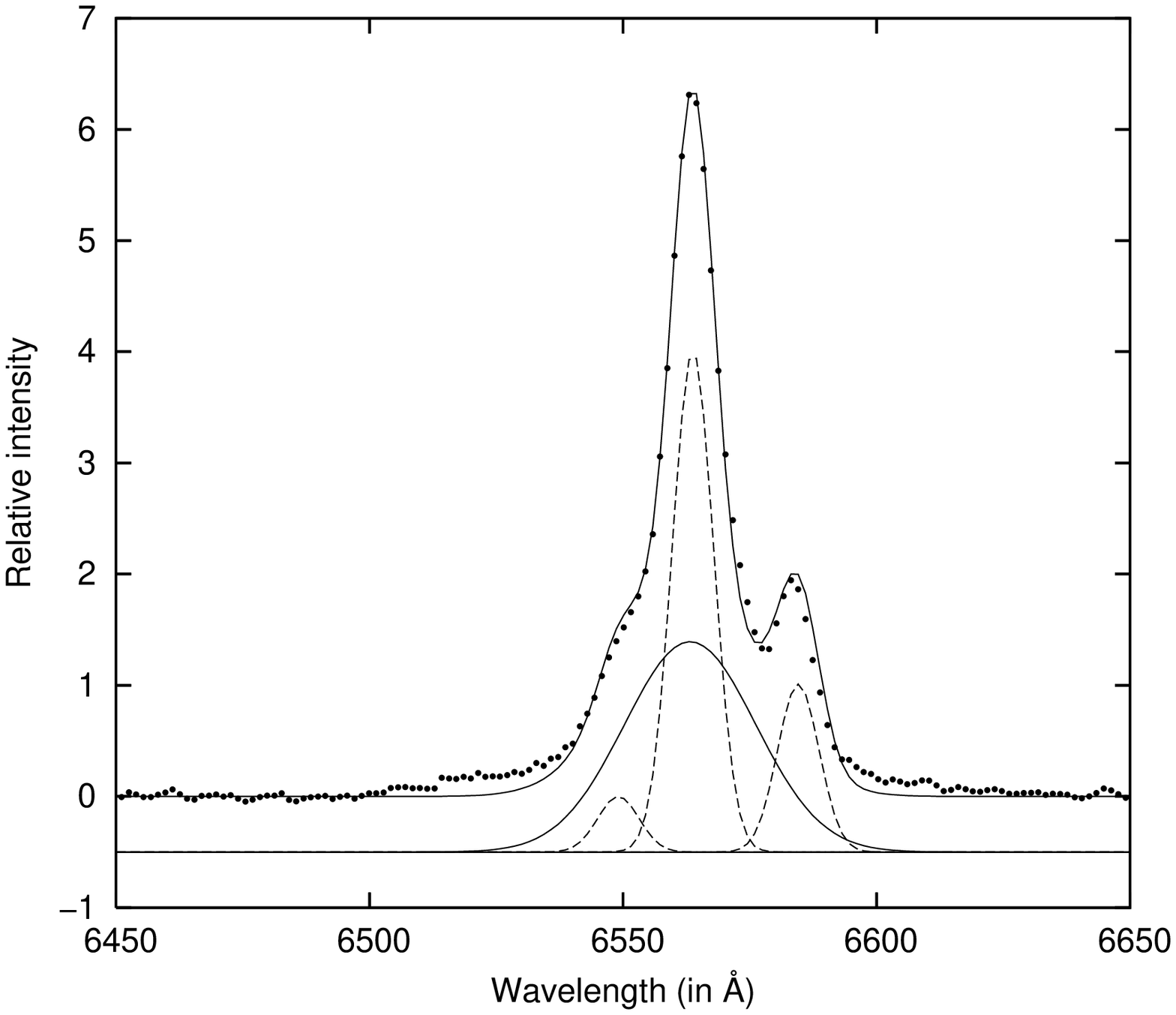}
\caption{  The decomposition of the observed H$\alpha$+[NII] spectral
lines (dots) where H$\alpha$ is decomposed with three (up) and two (down)
Gaussian
functions.} \label{fig_06}
\end{center}
\end{figure}

\begin{figure}
\begin{center}
\includegraphics[width=0.45\textwidth]{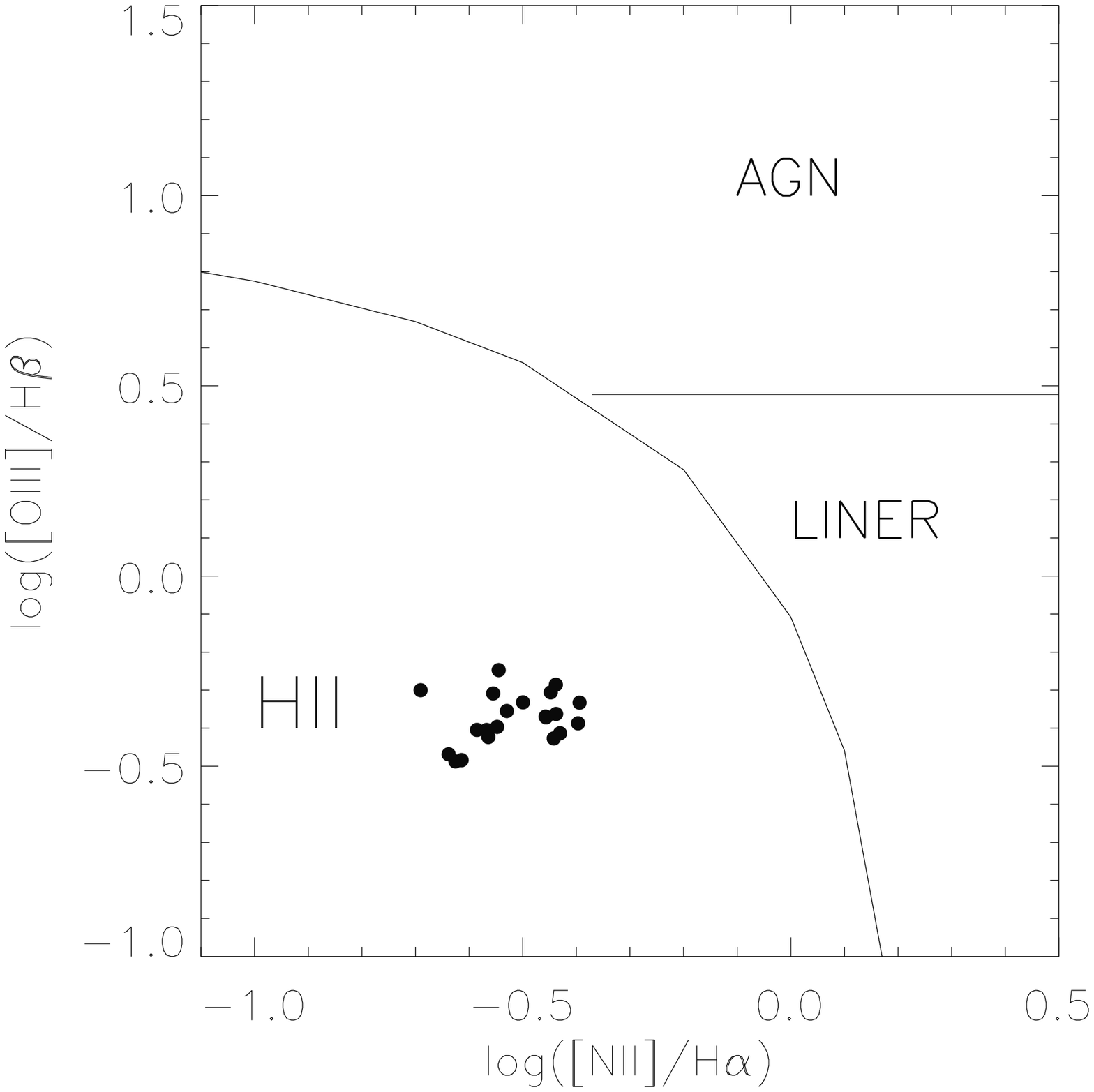}
\includegraphics[width=0.45\textwidth]{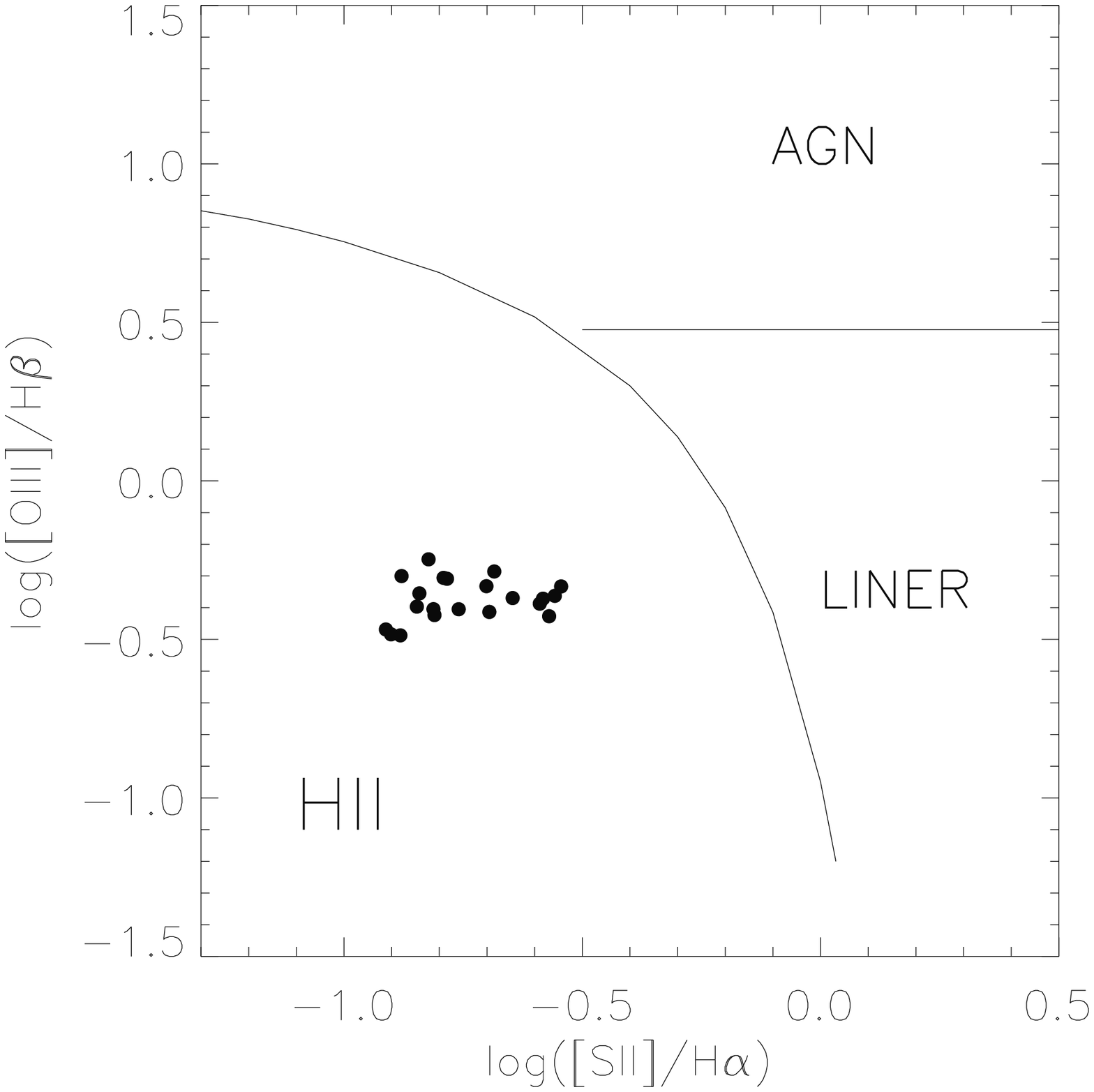}
\caption{{ The diagnostic diagrams of Mrk 493 nucleus. Only narrow Balmer line component is included. A dot presents the line flux ratio obtained from the same spaxel. }
}\label{fig_diagram}
\end{center}
\end{figure}

\begin{figure}
\begin{center}
\includegraphics[width=0.45\textwidth,angle=270]{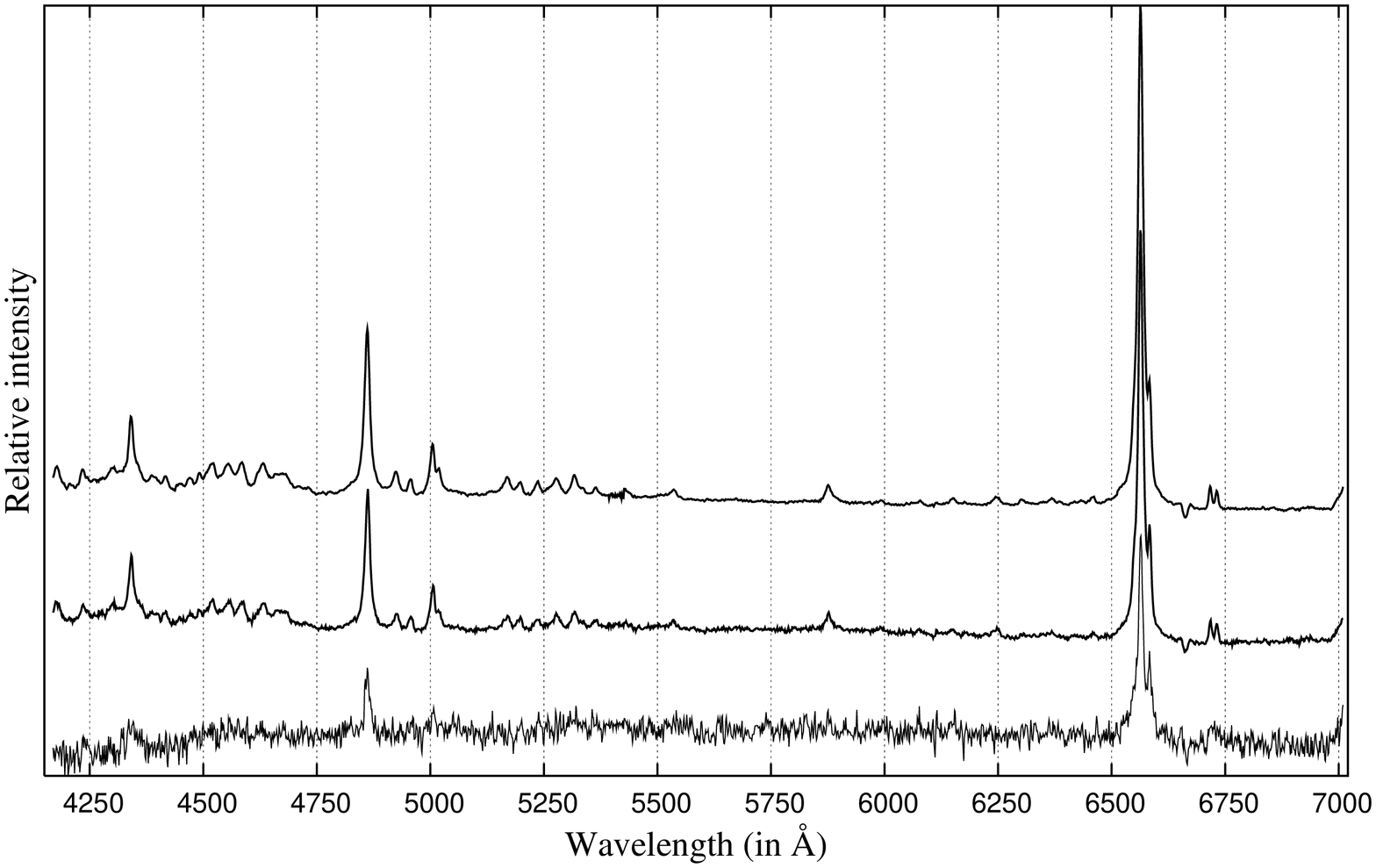}
\includegraphics[width=0.45\textwidth,angle=270]{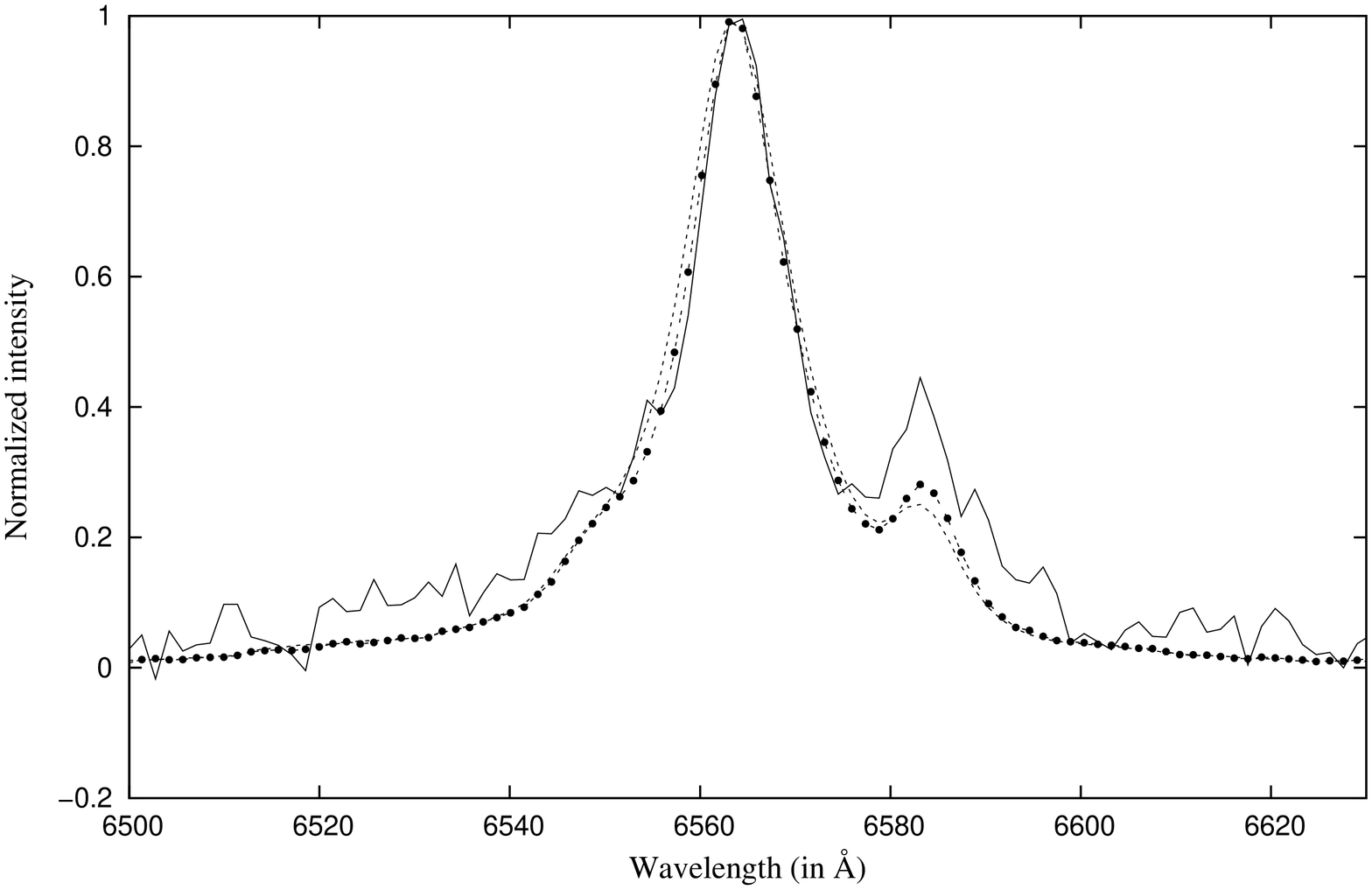}
\caption{{\it Top:} The spectrum from central spaxel (0,0) - up, from 2" from the center, (0,2) spaxel -middle and from  4" from the center, (0,4) spexel -down. The spectrum from spaxel (0,2) was multiplied  with factor 3 and spectrum from spaxel (0,4) with factor 27.
{\it Down:} The H$\alpha$ lines are normalized to one from these three spectra:
central spexel (0,0) -- dashed line, spaxel (0,2) -- dashed-doted line, and spaxel (0.4) -- solid line.}
\label{fig_diagram}
\end{center}
\end{figure}

\begin{table}
\caption{Log of MPFS  observations.} \label{speclog}
\begin{tabular}{@{}llllll@{}}
\hline Date & T$_{exp}$ & Sp. Range & Sp. Res. & Seeing \\
& (sec) & (\AA) & (\AA)  &  (arcsec) & \\
\hline
2004 May  21 &  4800     &    4150 - 5650   &   4   & $1.5\pm0.2$ \\
2007 May  18 &  7200     &    4300 - 7200   &   8   & $1.7\pm0.2$ \\
\hline
\end{tabular}
\end{table}

\begin{table}
\caption{The  parameters of  the best fit for the [O III] $\lambda\lambda$4959, 5007 \AA \ lines and Fe II template ($\lambda\lambda$ 4400-5400 \AA) for each spaxel (16 spectra from 2004 observations). Used notation:
'position' -  the position of the spaxel in arc secunds, where (0,0) is the central part; w [O III] - width of the [OIII]; [O III]/H$\beta$ - the flux ratio of the [O III] and H$\beta$ total flux; w Fe II and sh Fe II - widths and shifths of Fe II lines, respectively;  Fe II$_{total}$/H$\beta$ - the flux ratio of the Fe II$_{total}$ and H$\beta$ (total flux)
} \label{tab2}
     \begin{tabular}{|c|c c c c c|}
\multicolumn{6}{c}{}\\

\hline \footnotesize {position}& \footnotesize {w [O III]}& \footnotesize { [O III]/H$\beta$ }&\footnotesize {w Fe II 
} &\footnotesize {sh Fe II}& \footnotesize {  Fe II$_{total}$/H$\beta$} \\
\hline
( 0, 0)   &  310 &  0.058& 780   & 79    & 0.56   \\
(+1, 0)   &  290 &  0.057& 720   & 83    & 0.48  \\
 (- 1, 0)  &  280 &  0.061& 840   & 24    & 0.69  \\
 ( 0, +1)  &  270 &  0.045& 770   & 48   &  0.48  \\
( 0, -1)  &  270 &  0.051& 840   & 50   &  0.60 \\
(+1, -1)  &  260 &  0.055& 870   & 18    & 0.59  \\ 
(+1, +1)  &  270 &  0.060& 810   & 67    & 0.49   \\
(- 1, +1) &  250 &  0.054& 720   & 80    & 0.48   \\
(- 1, -1) &  250 &  0.048& 720   & 48    & 0.59  \\
( 0, -2)  &  290 & 0.069 &720    & 40    & 0.61    \\
(+1, -2)  &  260 & 0.061 &720    & 78    & 0.46   \\ 
(- 1, -2) &  190 & 0.045 &720    & 23    & 0.48   \\
( 0, +2) &  230 &  0.058& 720   & 50    & 0.40   \\
(+1, +2)  &  240 &  0.071& 720   & 96   &  0.49  \\
 (+2, 0)  &  200 &  0.059& 750   & -52   &  0.46\\ 
(+2, -1)  &  170 &  0.051& 750   & -67   &  0.47   \\
 
   \hline            
 average val.& 250$\pm$40& 0.056$\pm$0.007&760$\pm$50&42$\pm$46&0.52$\pm$0.08\\  
 \hline
      
   \end{tabular} 
\end{table}

\begin{table}
\caption{The same as in Table 2, but for the H$\beta$ BLR, ILR and NLR components. The width and shift of the H$\beta$ NLR component is assumed to be the same as one of the [OIII] lines.
} 
     \begin{tabular}{|c|c c c c c c c|}
\multicolumn{8}{c}{}\\          
  
\hline \footnotesize {position}& \footnotesize {w BLR H$\beta$}&\footnotesize {w ILR H$\beta$ }&\footnotesize {sh BLR H$\beta$ }&\footnotesize {sh ILR H$\beta$ }& \footnotesize {H$\beta$ BLR/H$\beta$ }& \footnotesize {H$\beta$ ILR/H$\beta$}&\footnotesize {H$\beta$ NLR/H$\beta$} \\
\hline    

 ( 0, 0)   & 2490  & 840  &-267  &205  & 0.35 & 0.35 &  0.30        \\  
 (+1, 0)   & 2490  & 820  &-633  &191  & 0.33 & 0.40 &  0.27       \\  
  (- 1, 0)  & 2490  & 900  &-294  &100  & 0.31 & 0.34 &  0.35      \\ 
  ( 0, +1)  & 2820  & 810  &-647  &184  & 0.38 & 0.35 &  0.27    \\         
( 0, -1)  & 2440  & 670  &-631  &171  & 0.34 & 0.44 &  0.22       \\          
(+1, -1)  & 2480  & 860  &-277  &172  & 0.31 & 0.36 &  0.33     \\          
(+1, +1)  & 2490  & 940  &-170  &176  & 0.30 & 0.39 &  0.32     \\   
 (- 1, +1) & 2400  & 800  &-443  &95   &0.43  &0.29  & 0.29       \\                          
                              (- 1, -1) & 2400  & 800  &-434  &209  & 0.38 & 0.25 &  0.37     
\\                                                       ( 0, -2)  & 2400  & 840  &-423  &115  & 0.40 & 0.29 &  0.30      \\         
(+1, -2)  & 2400  & 810  &-410  &188  & 0.43 & 0.26 &  0.30      \\         
 (- 1, -2) & 2100  & 690  &-424  &114  & 0.50 & 0.27 &  0.22    \\                            
                              ( 0, +2) & 2400  & 800  &-432  &105  & 0.48 & 0.32 &  0.21      
\\          
(+1, +2)  & 2100  & 750  &-397  &141  & 0.37 & 0.38 &  0.25      \\       
   (+2, 0)  & 2100  & 670  &-418  &242  & 0.40 & 0.22 &  0.38     \\                          
                             (+2, -1)  & 2100  & 690  &-444  &93   &0.49  &0.23  & 0.28       
 \\       
                                                                          
        \hline
 average val.& 2380$\pm$200 & 790$\pm$80&-421$\pm$133&156$\pm$47 &0.39$\pm$0.06&0.32$\pm$0.06&0.29$\pm$0.05\\
 \hline

   \end{tabular} 
\end{table}

       \begin{table}
\caption{ The same as in Table 2, but for  21 spectra (observed in 2007) of the H$\alpha$ BLR, ILR and NLR components. 
} 

     \begin{tabular}{|c|c c c c c c c c|}
\multicolumn{9}{c}{}\\          

\hline \footnotesize {position}& \footnotesize {w BLR H$\alpha$}& \footnotesize {w ILR H$\alpha$}&\footnotesize {w NLR H$\alpha$}& \footnotesize {sh BLR H$\alpha$}& \footnotesize {sh ILR H$\alpha$}&$\frac{\mathrm{H\alpha BLR}}{\mathrm{H\alpha}}$&$\frac{\mathrm{H\alpha ILR}}{\mathrm{H\alpha}}$&$\frac{\mathrm{H\alpha NLR}}{\mathrm{H\alpha}}$ \\
\hline       
 \footnotesize {( 0, 0)}  &  2460 &    700 &    300 &    -196  &   -16&    0.24&  0.35 & 0.41     \\ 
 \footnotesize {(+1, 0)}  &  1980 &    750 &    300 &    -352  &   13 &   0.18&  0.42 & 0.39   \\  
 \footnotesize {(- 1, 0)} &  1980 &    750 &    300 &    -359  &   14 &   0.23&  0.32 & 0.45    \\  
 \footnotesize {( 0, +1) }&  1980 &    690 &    300 &    -265  &   4  &  0.22&  0.37 & 0.41      \\ 
 \footnotesize {( 0, -1) }&  1740 &    640 &    310 &    -216  &   9  &  0.24&  0.34 & 0.42     \\  
 \footnotesize {(- 1, +1)}&  2040 &    700 &    290 &    -184  &   5  &  0.25&  0.32 & 0.43  \\  
 \footnotesize {(+1, +1) }&  2040 &    650 &    280 &    -169  &   13 &   0.27&  0.36 & 0.37   \\ 
\footnotesize { (- 1, -1)}&  2040 &    680 &    270 &    -181  &   14 &   0.26&  0.33 & 0.42   \\
 \footnotesize {(+1, -1) }&  2040 &    640 &    280 &    -139  &   10 &   0.26&  0.34 & 0.40    \\
 \footnotesize {(- 1, +2)}&  2040 &    820 &    300 &    -143  &   -90&    0.22&  0.23 & 0.55   \\ 
 \footnotesize {( 0, +2) }&  1770 &    700 &    290 &    -129  &   -15&    0.23&  0.29 & 0.48    \\
 \footnotesize {(+1, +2) }&  1770 &    650 &    270 &    -121  &   22 &   0.30&  0.28 & 0.42    \\
\footnotesize { (- 2, +1)}&  1770 &    560 &    300 &    -135  &   -24&    0.23&  0.24 & 0.54    \\
 \footnotesize {(+2, +1)} &  1770 &    640 &    250 &    -139  &   25 &   0.29&  0.34 & 0.37  \\ 
 \footnotesize {(- 2, 0)} &  1560 &    580 &    250 &    -142  &   16 &   0.29&  0.36 & 0.35      \\
 \footnotesize {(+2, 0)}  &  1560 &    580 &    230 &    -138  &   26 &   0.25&  0.42 & 0.33   \\  
 \footnotesize {(- 2, -1)}&  1560  &  630 &    250 &    -128  &   33 &   0.27&  0.34 & 0.39      \\
 \footnotesize {(+2, -1) }&  1560 &    600 &    260 &    -136  &   24 &   0.33&  0.28 & 0.40      \\ 
 \footnotesize {(- 1, -2)}& 1530  &   610  &   220  &   -139   &  51  &  0.29&  0.37 & 0.35     \\ 
 \footnotesize {( 0, -2) } & 1530  &   570  &   250  &   -137   &  46  &  0.29&  0.34 & 0.37       \\
 \footnotesize {(+1, -2) } & 1530  &   560  &   230  &   -126   &  31  &  0.31 & 0.40  &0.29      \\
        \hline
 \footnotesize {av. val.}&\footnotesize {1820$\pm$250}&\footnotesize {650$\pm$70}&\footnotesize {270$\pm$30}&\footnotesize {-175$\pm$70}&\footnotesize {10$\pm$30}&\footnotesize { 0.26$\pm$0.03}&\footnotesize {0.33$\pm$0.05}&\footnotesize {0.41$\pm$0.06}\\
 \hline

   \end{tabular} 
\end{table}    


\begin{thebibliography}{}

\bibitem{}
Afanasiev, V. L., Dodonov, S. N., Moiseev, A. V., 2001, in Ossipkov L. L.,
Nikiforov I. I., eds, Proc. Int. Conf. Stellar Dynamics: from Classic to
Modern, Sobolev Astron. Inst. St Petersburg, p. 103

\bibitem{}
Afanasiev, V.L., Moiseev, A.V., 2005, Astronomy Letters, 31, 194
[astro-ph/0502095]

\bibitem {}
Arribas, S.,  Mediavilla, E.,
Garc\'{\i}a-Lorenzo B.,  del Burgo C., and Fuensalida J.J., 1999, A\&AS, 136, 189

\bibitem {}
Boller, T., Brandt, W. N., Fink, H.  1996, A\&A, 305, 53

\bibitem{}   Boroson, T. A., Green, R. F. 1992, ApJS, 80, 109

\bibitem{} Botte, V., Ciroi, S., Rafanelli, P., Di Mille, F. 2004, AJ, 127,
3168

\bibitem{} Buta,  R.,  Combes, F., 1996, Fundam. Cosm. Phys., 17, 95

\bibitem {}
Canalizo, G., Stockton, A., Roth, K. 1998, AJ, 115, 890

\bibitem{} Collin, S., Joly, M. 2000, NewA Rev., 44, 531

\bibitem{} Crenshaw, D. M., Peterson, B. M., Korista, K. T., Wagner, R.
M., Aufdenberg, J. P. 1991, AJ, 101, 1202

\bibitem{} Davies, R.I., Tacconi, L.J., Genzel, R. 2004, ApJ, 613, 781

 \bibitem{} Davis, R.I., Mueller S\'anchez, F., Genzel, R., Tacconi, L.J.,
Hicks, E.K.S.,
Friedrich, S. 2007, ApJ, 671, 1388.

\bibitem{deo06}
Deo, R.P., Crenshaw, D.M., Kraemer, S.B, 2006, AJ, 132, 321

\bibitem{}
Dimitrijevi\'c, M. S., Popovi\'c, L. \v C., Kovacevi\'c, J., Da\v
ci\'c, M., Ili\'c, D. 2007, MNRAS, 374, 1181

\bibitem {}
Gliozzi, M., Papadakis, I. E., Brinkmann, W. P. 2007, ApJ, 656,

\bibitem{} Hamann, F.,  Ferland, G. 1999, ARA\&A, 37, 487

\bibitem {Hu08a}
 Hu, C., Wang, J. M., Chen, Y. M., Bian W. H., Xue S. J., 2008, ApJL,683, 115


{ 
\bibitem {Hu08b} Hu, C. Wang, J.-M. Ho, L. C., Chen, Y.-M., Zhang, H.-T., Bian, W.-H., Xue, S,-J. 2008, ApJ, 687, 78}

\bibitem{} Joly, M. 1993, Ann. Phys. Fr., 18, 241

\bibitem{} Jogee, S., 2006,  Lecture Notes in Physics,  693, 143

\bibitem[\protect\citeauthoryear{Kennicutt}{1998}]{ken98}
Kennicutt, R.C., 1998, ARA\&A, 36, 189

\bibitem[\protect\citeauthoryear{Klimek et al.}{2004}]{kli04}
Klimek, E.S., Gaskell, M. C., Hedrck, C. H. 2004, ApJ, 609, 69.

\bibitem{} Komossa, S. 2008, Rev. Mex. AA C, 32, 86

\bibitem{kov08}
Kova\v cevi\'c, J., Popovi\'c, L. \v C., Dimitrijevi\'c, M. S. 2008, in preparation.

{
\bibitem{Ku08} Kuehn, C. A., Baldwin, J. A., Peterson, B. M., Korista, K. T. 2008, ApJ, 673, 69}

\bibitem {} Lawrence, A.,  Elvis, M., Wilkes, B., McHardy, L., Brandt, N.
1997, MNRAS, 285, 879

\bibitem {}
Leighly, K. M. 1999a, ApJS, 125, 317

\bibitem {}
Leighly, K. M. 1999b, ApJS, 125, 297

\bibitem {}
L\'ipari, S.L., Terlevich, R.J. 2006, MNRAS, 368, 1001.

{ 
\bibitem{} Lipari, S., Colina, L., Macchetto, F. 1994, ApJ, 427, 174}

\bibitem{} Moiseev A.V., Vald{\'e}s J.R., Chavushyan V.H., 2004, A\&A,
421, 433

\bibitem{} Mathur, S. 2000 NewAR, 44, 469

\bibitem{munoz07}
Mu\~{n}oz-Mar\'{i}ne V.M.,  Gonz\'{a}lez Delgado R.M., Schmitt H.R., Cid
Fernandes R., et al., 2007, AJ, 134, 648

\bibitem {} Nagao, T., Murayama, T., Taniguchi, Y. 2001, ApJ, 546, 744

\bibitem{}  Netzer, H. 2006, in `Physics of Active Galactic Nuclei at all
Scales', eds. by Alloin D.,
Johnson R.  and Lira P., Lecture Notes in Physics,  693, 1


\bibitem {} Osterbrock, D. E., Pogge, R. W. 1985, ApJ, 297, 166


\bibitem{} Popovi\'c, L. \v C., Mediavilla, E., Bon, E., Ili\'c, D. 2004,
A\&A, 423, 909.

\bibitem{}  Sigut, T.A.A. \& Pradhan, A.K. 2003, ApJS, 145, 15

\bibitem{} Sulentic, J. W.,  Marziani, P., \& Dultzin-Hacyan, D. 2000, ARA\&A,
38, 521

\bibitem {}
Wang, J.-M., Zhang, E.-P. 2007, ApJ, 660, 1072


\bibitem{} Veilleux, S., Osterbrock, D. M., 1987, ApJS, 63, 295

\bibitem{}
Veron-Cetty, M.-P., Joly, M., Veron, P., Boroson, T., Lipari, S.,
Ogle, P. 2006, A\&A,451. 851

\end{thebibliography}
        \end{document}